\documentclass[preprint,showpacs,preprintnumbers,amsmath,amssymb]{revtex4}

\begin{document}
\baselineskip=14pt
\preprint{nucl-ex/0203012}
\input{psfig.sty}
\title{Calculation of the Ultracold Neutron
Upscattering Loss Probability in Fluid Walled
Storage Bottles using Experimental 
Measurements of the Liquid Thermomechanical Properties
of Fomblin}

\author{S.K. Lamoreaux}

\affiliation{University of California\\
 Los Alamos National
Laboratory, Physics Division P-23\\ 
Los Alamos, New Mexico 87545 USA}

\author{R. Golub}

\affiliation{ Hahn-Meitner Institut, Gleinicker Str. 100\\
 D-14109 Berlin, Germany}

\date{\today}

\begin{abstract}
Presently, the most accurate value of the free neutron
beta-decay lifetime result from measurements using fluid-coated ultacold
neutron (UCN) storage  bottles.  The purpose of this work is to
investigate the temperature dependent UCN loss rate from
these storage systems. 

To verify that the surface properites of Fomblin films
are the same as the bulk properties,
we present experimental measurements of the properties of
a liquid ``Fomblin'' surface obtained by the quasielastic
scattering of laser light.  The properties include the surface
tension and viscosity as functions
of temperature.  The results are compared to measurements
of the bulk fluid properties.  

We then calculate the upscattering rate of 
UCN from thermally excited surface capillary waves on
the liquid surface and compare the results to experimental
measurements of the UCN lifetime in Fomblin fluid-walled
UCN storage bottles, and show that
the excess storage loss rate for UCN energies near
the Fomblin potential can be 
explained.  The rapid temperature dependence of the
Fomblin storage lifetime is  explained by our analysis.
 
\end{abstract}


\maketitle

\section{Introduction}

The concept of a fluid-walled ultracold neutron (UCN) storage bottle
is due to C. Bates \cite{bates}  who developed and
demonstrated the technique at the Risley reactor
in the U.K.  The basic idea is that a hydrogen-free
fluorinated oil (Fomblin) wets glass and metal surfaces, and thereby
produces a clean, chemically perfect, microscopically
flat and reproducible UCN reflecting surface.  
Fomblin is quite viscous at room temperature; when it is sprayed
onto surfaces, it drains away leaving a thin stable film.  The
fluid readily seals gaps and holes associated with UCN
entrance and exit valves, provided that the mechanical parts
fit rather closely.  Fomblin also has a very low vapor pressure.
Some of the most accurate determinations of the neutron
$\beta-$decay lifetime have been obtained with Fomblin-walled
storage cells. \cite{mampe,arzu}

There is some uncertainty in the chemical formula of Fomblin
Y Vac 18/8 (manufactured by Ausimont/Montedison Group), the
material most widely used in UCN storage experiments.  
The mean molecular weight is 2650, and the stoichiometry
is roughly $\rm C_{3}F_{6}O$, with density at
20$^\circ$ C of 1.89 g/cc, with a mean effective
UCN potential of $V=106.5$ neV.  
The UCN loss coefficient (\cite{ucnbook}, Eq. (2.66)) is given by
the ratio of the imaginary $W$ to real part $V$ of the effective
UCN potential,
\begin{equation}\nonumber
f={W\over V}={\sigma_l k\over 4 \pi a}
\end{equation}
where $\sigma_l$ is the total $1/v$ loss cross section for
neutrons with wavenumber $k$, and $a$ is the coherent
scattering length.  $f$ has contributions from
nuclear absorption and bulk process, e.g.  upscattering which
occurs for temperature $T>0$.  
For the range of formulae
for Fomblin given in \cite{fomform}, the possible values of
$f$ due to nuclear absorption lie in the range
\begin{equation}\label{f}
2.5\times 10^{-7}<f<6\times 10^{-7}.
\end{equation}
The uncertainty in the Fomblin molecular
formula given in \cite{fomform}  is a result of
the coherent scattering lengths of 
C, F, and O being quite similar (6.65, 5.60, 5.81 fm, respectively)
so the effective UCN potential $V$ is relatively insensitive
to the relative numbers of each atomic species.  On the other hand, 
the nuclear absorption cross sections of C, F, and O
are quite different
(3.5, 9.6$\pm$.5, 0.19 mb, respectively, for thermal (2200 m/s) neutrons).
Thus, the stoichiometry of Fomblin could be better determined
by simultaneously considering $f$ extrapolated to $T=0$, 
which is quite sensitive to
the atomic fraction of F because it has a comparatively larger absorption
cross section, and $V$ which cannot be used to
effectively determine the relative amounts of F and O.

For experiments operated near room temperature
(290 K - 320 K), it is
found that 
\begin{equation}
2\times 10^{-5}<f<6\times 10^{-5}
\end{equation}
and the
temperature dependence of $f$ is much faster than being
proportional to  $T$ or $\sqrt{T}$ as might be
expected due to atomic and molecular fluctuations in
the wall (see, e.g., \cite{ucnbook}, Sec. 2.4.6) , 
and if fact seems to
follow the viscosity temperature dependence
rather closely.  Some attempts were
made to explain the observed temperature dependence of
$f$.  For example,
if there is a $1/v$ cross section
for slow neutrons due to upscattering in
the fluid, this cross section can be used to modify
$W$ to include this loss mechanism and in this regard,
the transmission of 60 \AA\space neutrons
through a sample of Fomblin as a function of sample temperature
was measured \cite{ruddies}.  
These measurements are discussed in \cite{mampe} where
it is stated that the temperature dependence of $f$
comes within a factor of 1.5 of explaining the experimentally
observed temperature dependence of the loss rate.  
The loss rate for the lowest temperature 218 K \cite{moroz}
implies  $f=6\pm 2\times 10^{-6}$
which is still an order of magnitude  greater than Eq. (\ref{f}).
More recent measurements by by Morozov et al.
of the transmission of 9 m/s 
neutrons through a 4 mm thick sample as a function
of temperature are consistent with these results
\cite{moroz2}.

It is possible that $f(T)$ has contributions from different
sources.  The purpose of our work reported here is to evaluate
the upscattering loss of UCN due to thermally excited surface
capillary waves.  Pokotilovski has investigated the energy
broadening of (monochromatic) UCN stored in a fluid walled 
bottle \cite{poko1,poko2};
this is a question closely related to the upscattering loss,
but it seems to us that the temperature dependent
loss rate is an experimentally  better studied
phenomenon but is lacking a fundamental explanation.
The principal data we will address were
obtained by Richardson et al. \cite{dave}.

\section{Experimental Determination of the Thermomechanical
Properties of Fomblin Y Vac 18/8}

Although the properties of this material are available from
the manufacturer, we thought it prudent to perform measurements
ourselves under conditions similar to those used in an
experiment.  The sample of Fomblin Y Vac 18/8 (hereafter
referred to as Fomblin) was obtained
from that used in \cite{mampe}.  These is anecdotal evidence
that the viscosity changes for degassed samples in vacuum
compared to samples that have been stored a long time in air.
We therefore performed our measurements with the Fomblin under vacuum
when possible, with the material being introduced into the
various apparatuses to be discussed
by distillation under high vacuum ($P<5\times 10^{-7}$ torr).

These measurements were done in Aug. 1991; some details 
were lost (e.g., specific equipment models employed) but given
the modest accuracy of the results, these details are
not important.

\subsection{Mechanical Viscosity Measurement}

The viscosity was determined by the time it took for a
known volume of Fomblin to drain through a Pyrex capillary of
known length and diameter, under the influence of gravity.
The glass system comprising two relatively large volumes connected
by a capillary is shown in Fig. 1.  Marks were placed on
the large tubes corresponding to a volume $v=0.425 \pm .005$ cm$^3$.
The capillary diameter was $2.00\pm 0.01$ mm, with length
4.8$\pm 0.1$ cm.  

Fomblin was distilled into the glass measurement apparatus 
under high vacuum; the glass was sealed off from the vacuum
system, and the time to drain the known volume was measured
as a function of temperature, with the temperature determined
by a water bath.  The kinematic viscosity is given by (\cite{ll},
Eq. (17.10))
\begin{equation}
\nu={\pi g\over 8 (V/t)}R^4
\end{equation} 
where $\nu$ is the kinematic viscosity, $g$ the acceleration of
gravity, $V$ is the known volume, $t$ is the time to drain the
known volume, and $R$ is the capillary radius.  The time to
drain the volume varied from about 150 sec at the lowest temperatures,
to about 2 sec at the highest temperatures; at the lowest
temperatures, the constancy and value of the temperature was
the major uncertainty ($\pm 1.2$ K), while at the high temperatures, the
timing (to 0.1 sec accuracy) was the principal limitation to
the accuracy.  In addition, there
is a small correction associated with
the change in height of the liquid layer
as the Fomblin drains, but it is less
than 5\%. The results are shown in Fig. 2 and are in
good agreement with, and follow the trends of, 
values reported by the manufacturer
(1.9, .09, and .02  cm$^2$/s at 293, 393, 493 K respectively).
In the range of measurements, the viscosity is adequately
described to within experimental error by the following empirical formula:
\begin{equation}
\nu(T)=2.0\times 10^7e^{-0.055T} +900T^{-3/2}
\end{equation}
where $\nu$ is in cm$^2$/s, and $T$ is measured in K, with estimated
accuracy over the temperature range 280-308 K which will be used
later.

\subsection{Thermal Expansion}

Fomblin was distilled into a Pyrex capillary tube of 2 mm inside
diameter, filling the tube to a length of about 5 cm.
The tube was sealed off from the vacuum system, and the 
relative change in length of the sample as a function of
temperature was measured.  The results are shown in Fig. 3;
the principal limits to accuracy were the length measurement ($\pm 0.2$ mm)
and temperature control ($\pm 0.5$ K).  The thermal expansion
of Fomblin is about 100 times larger than Pyrex, so the
expansion of the glass is an insignificant correction. The
density as a function of temperature is given by
\begin{equation}
\rho(T)=1.89[1+8.97\times 10^{-4}(294-T)]\  {\rm g/cm^3}
\end{equation}
where $T$ is measured in K, and the accuracy of this fit is
about 5\%.
The temperature dependence of the density is particularly
important because it leads directly to a temperature dependence
of the effective UCN potential.  This information was not
available from the manufacturer.

\subsection{Surface Tension}

The surface tension was measured using a Cenco student demonstration
tensiometer which in essence measures the restoring force against
a length of wire pulled up against the liquid surface.  Unfortunately, there
was no simple way to perform this measurement on a sample under
vacuum.  The apparatus was calibrated with deionized water.
For Fomblin, the surface tension was found to be
\begin{equation}
\alpha= 24\pm 1\  {\rm dyne/cm}
\end{equation}
and was independent of temperature to within measurement accuracy.
The value quoted by the manufacturer is 20 dyne/cm.

\subsection{Liquid Properties by Light Scattering}

By use of a heterodyne laser scattering technique, the dynamic liquid
surface properties of Fomblin could be measured directly
without physically contacting the surface in any significant
way.  The idea is that the surface undergoes a sort of brownian
motion due to thermal fluctuations.

The theory describing the two-dimensional thermally excited
surface capillary waves has been developed by Bouchiat and
Meunier; when presented as a time and two-dimensional surface
Fourier transforms, the mean-square height fluctuations are described by
\cite{bom}
\begin{equation}\label{bm}
P_{\bf q}(\omega)=P(\omega,q)={k_B T\over \pi\omega}
{q\over \rho}\tau^2 {\rm Im}\left[{1\over 
D(-i\omega \tau)}\right] 
\end{equation}
where 
\begin{equation}
D(S)=y+(1+S)^2-\sqrt{1+2S}
\end{equation}
\begin{equation}
y={\alpha \rho\over 4\eta^2 q},\ \ \ \tau={\rho\over 2\eta q^2}
\end{equation}
where, as before, $\alpha$ is the surface tension, $\rho$ is
the density, and $\eta$ is the viscosity $\eta=\nu\rho$ where
$\nu$ is the kinematic viscosity.

Directed laser 
light reflecting from this surface will be scattered by time-varying
surface disturbances, with surface wavevector $q$ and frequency
$\omega$.  By purposely aiming a fraction of the light specularly
reflected from the surface in an angle corresponding to diffraction
by a disturbance with wavenumber $q$ (the ``reference beam'') to the detector, 
the time-dependent scattered
component can be measured by beating with the reference beam.  
The reference beam is produced by placing a weak transmission
diffraction grating in the specularly reflected beam (this grating
also intercepts the scattered light which is extremely weak).
This technique was proposed and developed by H\aa rd et al.
\cite{hard}  The specific apparatus that we used is
described in \cite{ajp}, with the exception of the vacuum
sample cell.  As discussed in \cite{ajp}, data was aquired with
the apparatus on a vibration isolated optical table.  In addition,
the vacuum sample cell isolated the liquid surface from disturbing
air currents which were a problem for free liquid surfaces in air.

The Fomblin sample was contained under vacuum in a cubical cell;
the cell was constructed of 5 mm thick float glass plates, and had internal
dimension of 4 cm$^3$, glued together with Torr-Seal Epoxy (Varian).
A 5 mm diameter Pyrex tube was glued into a hole drilled in the top plate;
Fomblin was distilled into the cell (to a depth of 3 mm) through
this tube and then sealed off from the vacuum system.

The cube was cooled/heated from the bottom with a Peltier device, and
the temperature was controlled with a simple feedback loop.  Light
was directed onto the liquid surface at an angle of approximately 70$^\circ$
from normal through one of the vertical walls of the cell; the scattered
and specularly reflected light exited the opposite vertical wall.
The laser light was detected with a PIN photodiode, and the
heterodyne signal was Fourier analyzed with a Hewlett-Packard audio
frequency spectrum analyzer.  Results as a function of temperature
and $q$ are shown in Fig. 4.  Also shown are four-parameter fits
to Eq. (\ref{bm}); these parameters include $\tau$, $y$, an amplitude
factor, and a DC offset.  Results of the fits are given in Table I.

The inferred values of $\nu$ and $\alpha$ are also given in
in Table I.  As can be seen, these results agree with previous
mechanical measurements; the values of $\nu$ are
plotted in Fig. 2, shown as squares.

\section{Calculation of the Upscattering Loss
Rate due to Thermally-Excited 
Surface Capillary Waves}

The surface waves described in Sec. II D
can inelastically scatter (diffract) a UCN
that reflects from the surface.  The probability to upscatter
a UCN to an energy outside the UCN range can be calculated
using Eq. (\ref{bm}) when it is properly normalized.
The formalism describing non-specular inelastic scattering was
developed by Pokotilovski \cite{poko2} and we will
use his results with minor modifications.

\subsection{Normalization of Eq. (\ref{bm})}

The work of \cite{bom} was intended to derive the spectrum
of the surface fluctuations; the absolute normalization
appears to have been of secondary importance.  For our case,
the overall magnitude is of crucial importance.

First, the static average surface distortion, based on
statistical mechanics considerations, is given as
\begin{equation}\label{static}
\overline{|\zeta_{\bf q}(0)|^2}={k_BT\over \alpha q^2 {\cal A}}
\end{equation}
where $\cal A$ is the surface area.  If we imagine $\cal A$ to
be a square with dimensions $L$ so ${\cal A}=L^2$, the effective
spread in $q_x$ and $q_y$ is
\begin{equation}
\Delta q_{x}=\Delta q_y
={2\pi\over L}\rightarrow {1\over {\cal A}}={1\over (2\pi)^2}
\ dq_x\ dq_y  
\end{equation}
and therefore, to cast Eq. (\ref{bm}) per unit $q_x q_y$, a factor
of $1/(2\pi)^2$ is required.

Second, the integral of $P(\omega,q)$ over $\omega$ should
give the static result, Eq. (\ref{static}):
\begin{equation}
\int_{-\infty}^{\infty} P(\omega,q)d\omega={k_BT\over \pi}
{q\over \rho} \tau^2{\rm Im}\int_{-\infty}^{\infty} {dx\over x}\left[{
1\over D(-ix)}\right]={k_B T\over \pi}{q\over \rho}\tau^2{\pi\over y}
={k_B T\over \alpha q^2}
\end{equation}
where the integration was performed by noting that the only
pole of the integrand is at $x=0$; by integrating along
the real axis and taking a semicircle around the origin,
the integral is $-i\pi$ times the residue at $x=0$ which
is $1/(-iy)$.

\subsection{Possible Corrections to Eq. (\ref{bm})}

The coating obtained by spraying Fomblin and allowing the
material to drain to a thin, stable film are of order $10^{-4}$ cm
thick or greater.  The multiplicative
correction factor to the capillary wave
dispersion relationship, due to finite film thickness,
is very roughly proportional to $\tanh(qd)$ where $d$ is
the film thickness (\cite{ll}, Sec. 62, Problem 1).  
For $q$ of interest for UCN
scattering, $10^{-4}{\rm cm}\times q>>1$ so $\tanh qd\approx 1$
 so we expect no significant
effect due to film thickness.  However, at higher temperatures
where the viscosity is low, the films might drain to layers
thin enough so that there is a substantial correction.  The
net effect will be to reduce the upscatter rate.

We are also considering rather high frequency waves;
it might be expected that the Fomblin has some frequency
dependence.  However, Fomblin appears to be a Newtonian
fluid (e.g., light scattering and mechanical measurements
had no anomalous behavior) and the Fomblin molecules are
not very large on the scale of what is considered a polymer.
The molecular collisional frequency of Fomblin molecules,
which is on of the factor that determine the viscosity,
is much higher than the frequencies of interest for UCN
scattering (e.g., $10^8$ Hz compared to more than $10^{10}$ Hz).  
For the analysis presented here, we will assume that the
low frequency measurements described earlier in this paper
are applicable.

\subsection{\bf Kinematics}

Let us consider an incident neutron with wavevector
(we set the azimuthal angle of the incoming wave to
zero without loss of generality); referring to Fig. 5,  
\begin{equation}
{\bf k_i}=k_{ix} \hat x + k_{iz}\hat z=k_i(\sin\theta_i
\hat x  + \cos\theta_i\ \hat z);\ \ 
\ k_i^2=2mE_i/\hbar^2
\end{equation}
where the polar angle $\theta_i$ is defined
relative to the surface normal which is taken along $\hat z$,
and $E_i$ is the incident UCN energy.
The surface wave disturbance can be described by
\begin{equation}
{\bf q}=q_x \hat x + q_y \hat y=q(\cos\phi_q \hat x + \sin\phi_q \hat y).
\end{equation}
where $\phi$ is the azimuthal angle relative to the incident
neutron momentum.

When a UCN diffracts from the time varying surface, energy and
momentum are conserved (we consider first upscattering, e.g.,
the UCN gains energy). 
\begin{equation}\label{econ}
k_f^2=k_i^2+ {2m\omega\over \hbar}.
\end{equation}

We can consider a
UCN of specified $\bf k_i$, hence $|\bf{k_i}|$ is specified, and choose
$\omega$ which determines $|\bf{k_f}|$; choosing $\bf q$ determines
$\theta_f$ and $\phi_f$, the polar angles of the outgoing neutron
wavevector.  The components of the wavevector in the liquid
surface plane (denoted by subscript {\bf s}) change on
reflection by
\begin{equation}
{\bf k_{fs}}={\bf k_{is}}+{\bf q}
\end{equation}
or
\begin{equation}
k_{fx}=k_i\sin\theta_i+q\cos\phi;\ \ \ \ \ k_{fy}=q\sin\phi.
\end{equation}
For the component of $\bf k_f$ normal to the surface,
\begin{equation}\label{constr}
k_{fz}^2=k_f^2-k_{fx}^2-k_{fy}^2\geq 0
\end{equation}
where the inequality constrains the values of $\bf q$
for allowed scattering. This sets a range on the magnitude
of $\bf q$:
\begin{equation}
0\leq q\leq q_{max}=k_f+k_i.
\end{equation}
In particular, we can define a kinematic factor such that
\begin{equation}\label{kink}
K(E_i,\theta_i,\omega,q,\phi)=
\left\{\begin{array}{ll}
1 &\mbox{if $k_{fz}^2\geq 0$}\\
0 & \mbox{otherwise,}
\end{array}
\right.
\end{equation}
where $k_{fz}^2$ is given by Eq. (\ref{constr}).

\subsection{ Loss Probability per Bounce}

The probability of loss per bounce was calculated by
Pokotilovski \cite{poko2}:
\begin{equation}\label{pokoup}
w_{q,\omega_q}=	8 k_{iz}k_{fz}P(\omega,q)\left\vert {k_{iz}-iK_{iz}\over
k_{fz}+iK_{fz}}\right\vert =8Fk_{iz}k_{fz}P(\omega,q)
\end{equation}
where
\begin{equation}\label{mom}
K_{i,f\ z}^2=k_c^2-k_{i,f\ z}^2
\end{equation}
with $k_c$ the critical momentum for UCN reflection.  
This equation results from determining
the scattered wavefunction amplitude then calculating the probability 
current normal to the surface.  Pokotilovski was interested
in below-barrier upscattering in which case the factor $F=1$.

In the case where the final UCN energy is higher than the 
wall potential, some care must be used in calculating
the upscatter loss probability.
The UCN will be lost from the system if the final neutron energy
is greater than the wall potential.  This loss has
two components:  1. The upscattered reflected wave will be
lost from the system, 2. The component of the 
scattered wave within the wall (film) is not an evanescent
wave, but is a propagating wave (e.g., $K_{fz}$ in Eq. (6) of \cite{poko2}
or Eq. (\ref{mom}) above
is imaginary).  We must therefore account for this losses for
the wave transmitted into the material.  Referring to Eqs. (A6.21) and (A6.22) of \cite{ucnbook},
the amplitudes of the reflected and transmitted upscattered waves
are equal.  The probability current
perpendicular to the liquid surface determines
the loss \cite{asty}, e.g., in Eq. \ref{pokoup} the factor $k_{fz}\rightarrow
k_{fz}+\sqrt{k_{fz}^2-k_c^2}$.  
Thus,
\begin{equation}\label{F}
F=\left\{ \begin{array}{ll}
1 &\mbox{if $k_{fz}^2<k_c^2$}\\ 
{k_c^2\over  k_{zf}[(k_{fz}^2-k_c^2)^{1/2}+k_{fz}]} & \mbox{otherwise.}\\
\end{array}
\right.
\end{equation}

The UCN will be lost from the system if its final energy
is greated than the wall potential $V$.
We can recast this in terms of allowed wave frequencies: For
UCN loss to occur,
\begin{equation}
\omega+\omega_0>\omega_c=V/\hbar;\ \ \ \ \hbar\omega_0=E_i 
\end{equation}
where $E_i$ is the incident UCN energy.  Thus, there is
a minimum $\omega$ for upscattering loss,
\begin{equation}
\omega_{min}=\omega_c-\omega_0.
\end{equation}

The differential probability that a UCN, with $E_i$, $\theta_i$, and
$\omega$ specified, is upscattered and lost from the
system due to surface waves in a small range around
a value $\bf q$ is given by
\begin{equation}
d\mu_{sw}=8Fk_{iz}k_{fz} w_i{P(\omega,q)\over (2\pi)^2} 
dq_x dq_y d\omega
\end{equation}
where $w_i$ is a weighting factor giving the probability for
a UCN to be incident at angle $\theta_i$,
\begin{equation}\label{weight}
w_i=2\sin\theta_i\cos\theta_i,
\end{equation}
and $dq_xdq_y=qdqd\phi$.

The total probability for loss is equal to the integral of
$d\mu$ over $q$ and $\phi$, followed
by an integration over  $\theta_i$, and finally over $\omega$
with  $\omega>\omega_{min}$,
\begin{equation}\label{loss}
\mu_{sw}(E_i,T,\omega_{min})=
\int_{\omega_{min}}^\infty\ d\omega\ \int_0^{\pi/2} w_id\theta_i
\left[2\int_0^\pi d\phi\int_0^{q_{max}}qdq\ K(E_i,\theta_i,\omega,q,\phi)
 8Fk_{iz}k_{fz}\ {P(\omega,q)\over (2\pi)^2}
\right]
\end{equation}
where $K$ is defined in Eq. (\ref{kink}), and the dependence of $\mu_{sw}$
on the temperature (due to, e.g., the viscosity
and density change of Fomblin) and incident UCN energy $E_i$ and
minimum energy $\omega_{min}$ to upscatter to greater
than $V$ are explicitly
indicated.
 
The integration is analytically unwieldy,
so it has been performed numerically.  The results, taking
into account the temperature dependence of the viscosity and
density (hence effective UCN potential)
of Fomblin are shown in Fig. 6.

The slow curvature with increasing temperature (hence decreasing
viscosity) results from the $y$ term in Eq. (\ref{bm}),
which is proportional to  the surface tension, becoming important.

\subsection{Reanalysis of Previous Data}

It is noted in \cite{dave} that a simple fit to a value of
$f$ and to losses through holes and gaps of total area $A_h$ does
not explain the observed lifetimes of a fluid-walled bottle for
energies within 10 neV of $V$.  There is enough flexibility in
$f$ and $A_h$ to adequately fit the lower values of energy;
as seen in Fig. 6, the surface wave upscatter loss probability
drops rapidly with energy.  However, the temperature variation
in $f$ as derived from these fits seems non-physical.

In general, a full monte-carlo analysis of the system is required.
For example, in a realistic storage system, the weight function
Eq. (\ref{weight}) becomes a function of geometry.  However,
80\% of the UCN loss for Eq. (\ref{loss}) occurs for angles
between 20$^\circ$ and 60$^\circ$, so at the 10-20\% level of
accuracy, we can neglect modification of the weight
function.   Furthermore, a full monte-carlo
analysis would require knowledge of the liquid surface specularity
for UCN reflection which is not known. Another fortunate feature
of the Ref. \cite{dave} measurements is that the cylindrical storage cell
height $2R=6.6$ cm leads to a small change in UCN energy as
a function of position in the cell (7\% effect) that can be
modelled, as described below, with 10\% accuracy.

For our reanalysis of the Ref. \cite{dave} experimental data, we
use oure experimentally determined values for viscosity (Eq. (4)),
density (Eq. (5)), and surface tension (Eq. (6)), and neglect the
uncertainties in these values.
We fitted the observed
lifetime vs. energy data to $f$ and $A_h$ was done as in \cite{dave}, but
included all data points and the surface wave upscatter loss
described by Eq. (\ref{loss}).  However, particularly for $E_i$
close to $V$, we need to take into account the effects of gravity
in the storage bottle which was a horizontal cylinder with
radius $R=3.3$ cm and  length $L=160$).  There are two
effects that need to be accounted for.
First, the upscatter loss function energy dependence is very 
steep for $E_i$ approching $V$.  Second, a UCN that upscatters
at the ``top'' of the storage vessel will gain energy as
it falls, and will eventually be lost from the system
if the final energy $E_f\geq V-2mgR$ \cite{asty}.  This effect can be
modelled by varying $\omega_{min}$ in Eq. (\ref{loss})
as a function of height in the bottle.  

The total average lifetime for a 
UCN with initial energy $E$ (monochromatic)
specified at a horizontal plane that 
intersects and divides the storage vessel into
two halves at the cylinder axis
can be readily determined.  The rate of loss can be written  
\begin{equation}
{1\over \tau(E)}={1\over \tau_\beta}+\ell^{-1} v(E)\bar\mu_{tot}(E)
\end{equation}
where $\tau_\beta$ is the neutron beta decay lifetime, $\ell$
is the mean free path,
\begin{equation}
\ell^{-1}={1\over 2L}+{1\over 2R}
\end{equation}
$v(E)$ is magnitude of the UCN velocity at the plane, and $\bar\mu_{tot}(E)$
is the total loss per bounce averaged over the storage
vessel walls.  The height-dependent total loss is given by  
\begin{equation}\label{lossave}
\mu_{tot}(E,\theta)= \left[1-{mgR\cos\theta\over E}\right]^{1/2}\left[
\mu(E-mgR\cos\theta)+\mu_{sw}(E-mgR\cos\theta,T
,\omega_{min}-2mgR\cos\theta) + {A_h\over 4 V_b}\right].
\end{equation}
where $\theta$ is the cylinder polar angle with $\theta=0$
being the bottom of the storage vessel, and
$\mu(E)$ is the $1/v$ material loss term  (\cite{ucnbook}, Eq. (2.70)),
\begin{equation}\label{mu}
\mu(E)=2f
\left[{V\over E}\arcsin\left({E\over V}
\right)^{1/2}-
\left({V\over E}-1\right)^{1/2}\right].
\end{equation}
As described in
the Introduction, $f$ has contributions from both nuclear absorption
and upscattering and represents the losses from the UCN
evanescent wave within the film during the time of reflection.  

The first factor in Eq. (\ref{lossave})
expresses the change in wall collision frequency
as a function of height and has a 2\% effect for $E\approx V$.
The average of the storage walls is given by
\begin{equation}
\bar\mu_{tot}(E)={1\over \pi}\int_0^\pi \mu_{tot}(E,\theta)d\theta.
\end{equation}
This average was numerically computed and then averaged over
with the
the energy resolution function given in \cite{dave}, Eq. (10),
to produce total loss curves as a function of temperature $T$;
$f$ and $A_h$ are left as fit parameters for a given $T$.

The most extensive data presented in \cite{dave} is for the
temperature $T=294$ K.  Including the surface wave upscattering
loss and fitting to the two parameters results in an excellent
fit, as shown in Fig. 7.  The decrease in $\chi^2$ (see Table II)
compared with
the fit results presented in \cite{dave} shows the importance
of the surface wave loss for UCN energies near the wall potential.
It should be noted that the excess loss is not associated with
slow heating of the UCN, as suggested in \cite{dave} but with a
direct upscattering loss.  As will be shown in the next section,
the heating/cooling effect probably does not alter the functional
form shown in Fig. 7.  

Fits to the 283 K and 308 K data are shown in Figs. 8 and 9, with
fit results tabulated in Table II.  Again, including the surface
wave upscattering loss provides an explanation of the extra loss
rate for the higher energy UCN.

It should be noted that the $\chi^2$ for these fits is extremely
sensitive to the energy, e.g., changing the energy for
either of the two highest points in Fig. 7 by $\pm 1$ neV 
changes $\chi^2$ by a factor of 0.5 to 2.  Therefore, the fact
that the spectrum shape can be varying during storage might
be evident in the data and analysis, but as can be seen by
directly considering Figs. 7-9, the simple analysis presented
here is adequate to within statistics, implying an accuracy
of 10-20\%.  Including the uncertainties in the measured
Fomblin properties would lead to a modest reduction in $\chi^2$. 

Finally, the values of $f$ for the various temperatures shown 
in Table II indicate a significant temperature dependence.  These
values, along with the value derived from \cite{moroz} discussed
in the Introduction are plotted in Fig. 10.  It can be seen
that a crude extrapolation to $T=0$ implies a reasonable range of values
for the residual (nuclear absorption) loss.

The observed variation of bottle lifetime as a function of
temperature is about 4 s/K \cite{mampe}; referring to
Fig. 7, the contribution to the loss rate
from Eq. (\ref{mu}) is about a factor of two
larger than the surface wave loss rate for 70 neV UCN.
Interpolating from Fig. 6, the change in $\mu_{sw}$
is about $5\times 10^{-8}/$K for 70 neV with the Fomblin
near room temperature, while from Table II $f$ varies as 
$df/dT= 2.3\pm 1 \times 10^{-7}/$K. 
The variation in $\tau$ with
change loss rate(s) can be estimated from Eqs. (27) and (30)
(the effect due to the change in the potential $V$ from thermal  expansion
is negligible)
\begin{equation}
\delta \tau/\delta T= \tau^2 \delta \mu_{tot}/\delta T =
\tau^2 {v\over \ell}\left[
1.44 \delta f + \delta \mu_{sw}\right]/\delta T
\end{equation}
which, for a 20 cm mean free path for 70 neV neutrons, with a storage lifetime
of 710 s \cite{mampe} 
implies
\begin{equation}
\delta\tau/\delta T= (710 {\rm\ s})^2 (1.44\times (2.3\pm 1)\times 10^{-7} 
+5\times 10^{-8}){\rm 370 \ cm/s\over 20\ cm}/\delta T
=
3.6\pm 1.6 {\rm\ s/K}
\end{equation}
which is in good agreement with 4 s/K.
The surface wave loss amounts to 20\% of the net total loss
for 70 neV UCN at $T=294$ K.  In \cite{mampe}, it is stated
that the observed upscatter loss in the transmission of 60\space\AA\space
neutrons comes within a factor of 1.5  of explaining the
observed loss.  Within the approximations here,
it would seem that the surface wave scattering can account
for most of the excess loss observed in this storage experiment.

\subsection{UCN Heating}

Heating of UCN occurs when energy is gained as before, but
when the final energy is still lower than the wall potential.
The calculation is carried out by integrating from near
$\omega=0$ to $\omega_{min}$ as defined in relation to Eq. (\ref{loss}).
The lower limit is set by the resolution of the measurement
apparatus, or, e.g., by the spectral width of the stored nearly
monochromatic UCN, taken here as $\pm 2$ neV, so $E_r=2$ neV 
and $\omega_r=E_r/\hbar$.  The heating probability
is thus obtained by modifying Eq. (\ref{loss}) (e.g., the
integration over $\omega$ is not  performed to get
$\mu_{sq}(E_i,T,\omega)$):
\begin{equation}\label{heat}
P_{heat}(E_i,T,\omega_r)=
\int_{\omega_r}^{\omega_{min}}\ \mu_{sw}(E_i,T,\omega) d\omega
\end{equation}
where $\omega_{min}$ was defined in relation to Eq. (\ref{loss})
and in this case represents the maximum energy change that
a UCN can receive and remain trapped.  The average energy
change is given by
\begin{equation}\label{upspec}
\Delta \bar{E}
(E_i,T,\omega_r)=\int_{\omega_r}^{\omega_{min}}\ \hbar
\omega \mu_{sw}(E_i,T,\omega) d\omega/P_{heat}(E_i,T,\omega_r).
\end{equation}

It is suggested in \cite{dave} that energy might be transferred
to the UCN at a rate of order $\delta E_{rms}=2(1.5)\times 10^{-11}$ eV
per collision.  The numerical results shown in Fig. 11 imply an energy
change of around $10^{-13}$ eV per collision, 
obtained by taking the upscatter
probability times the energy change.  Of course, very small
energy changes (less than the instrumental resolution) are
more probable, but many more reflections are needed to move a UCN from
the initial energy distribution; the $10^{-13}$ eV/collision
provides a reasonable estimate of the heating rate.  This result
is in agreement with the results of Pokotilovski \cite{poko1,poko2}
and we thus do not further elaborate the heating issue which is 
more fully discussed by him.

More recent measurements by Bondarenko et al. \cite{moroz3} indicate a heating
probabilities shown in Table III.  In this experiment,
UCN in a 0-52 neV energy range were stored in
a Fomblin-coated bottle, and upscattering to
energies between 52 and 106 neV were detected.
To calculate this probability, we assume an initial UCN energies
in the 0-52 neV range, and calculate the total probabilities
to scatter to the 52-106 neV range, using Eq. (\ref{heat}).
The energy-dependent upscatter probabilities are then averaged 
over the initial spectrum presented in \cite{moroz3}, Fig. 5.  
As presented in Table III, it can be seen that our results
are in reasonable  agreement with the experimental
results, to within the reported factor of 2 to 3 uncertainty associated
with neutron transport efficiency in the experimental apparatus.
A plot of the upscatter probability as a function of temperature,
for the parameters of this experiment, is shown in Fig. 12.

The cooling rate and average energy decrease can be calculated
in a similar fashion.  In this case, the integration limits
of Eq. (\ref{heat}) are 0 and $E_i/\hbar - \omega_r$.
The probability to upscatter or downscatter is shown in Fig. 13.
As discussed by Pokotilovski, the functions are very symmetric
about the incident UCN energy, so the effect of a small energy
change, either an increase or a decrease, tend to be averaged
away.  

The measurements by Bondarenko et al. \cite{moroz3}
also determined the probability for UCN with initial energy
around 13 neV, to downscatter to final energy in the range 0-11.5 neV,
and found a probability of about $10^{-6}$ per bounce.  The temperature
of the Fomblin was not specified; however, the measured
probability is consistent
with the results presented in Fig. 13 where the integrated probability
to downscatter from 10 neV initial energy to 
less than 8.5 neV is in the $10^{-6}$ 
range for both temperature
curves shown in Fig. 13.
(The variation in integrated probability with initial
energy is fairly slow; the calculated result would not be significantly
different for 13 neV initial energy.)  The rapid drop in
probablity with decreasing maximum final energy that is evident in Fig. 13
is consistent, within statisitics and uncertainties in
neutron transport, with the data
presented in \cite{moroz3}, Fig. 2.

\section{Conclusion}

As shown in Figs. 7-9, the UCN lifetime for energies near 100 neV
(e.g., the high energy data points presented in Ref. \cite{dave}) 
are well-described by upscattering loss due to surface waves.
In this analysis, the low-frequency kinematic viscosity and
surface tension were used.  The excellent agreement between
theory and experiment suggest that no modification of the low
frequency dispersion curve for the surface waves are required,
but we might anticipate that
for higher accuracy data, this simple analysis ultimately
might not be sufficient.

It was suggested in \cite{dave} that the excess high energy loss
is due to changes in the stored UCN spectral properties; 
the analysis presented
here indicates that direct upscattering to energies higher than
the wall potential fully account for the observed loss rate
beyond that due to nuclear absorption and upscattering
in the film (as parameterized by $f$ in Eq. (\ref{mu})) and losses
through holes and gaps.
The decrease in $\chi^2$ for
the re-analysis of the Richardson et al. data, with no
adjustable parameters, indicates that surface wave upscattering
laregely  describes the excess loss rate for UCN energy close
to the wall potential.
As described in Sec. F, and shown Figs. 11 and 12,
the heating and cooling probabilities are sufficiently  small so that, at
the level of accuracy of the data presented in \cite{dave}, the
effects of spectral evolution are not important.  However,
such effect are likely important for high-accuracy determination
of the neutron lifetime from Fomblin coated bottles.  

We have shown that the observed temperature dependence 
of the net loss rate has a significant component due to the
surface wave upscattering loss process.  In \cite{mampe},
the measured temperature dependence of the 
lifetime for 70 neV in a Fomblin coated bottle
is given as 4 s/K. This was not fully explained by the temperature
dependent $1/v$ cross section (e.g., the temperature
dependence of $f$ in Eq. (\ref{mu})).  The temperature
dependent loss implied by
measurements of transmission of 60\space\AA\space neutrons through Fomblin
was 33\% too small
\cite{mampe,ruddies}.  However,
from our analysis,
the surface wave upscattering accounts for up to 20\% of the
loss for 70 neV UCN.  This accounts for much of the reported discrepancy.
This, together with our estimate of the temperature variation
of $f$, explains
the UCN loss in Fomblin coated bottles to within experimental errors.
  
Recent Fomblin transmission  measurements using 9 m/s neutrons by
Morozov et al. \cite{moroz2} indicate that
$df/dT=9 \pm 5 \times 10^{-8}$/K which is in rough
agreement with the
result presented in Sec. IIE, based on Fig. 6.
Furthermore, the 
values of $f$ implied by the Ref. \cite{moroz2} measurements are
in agreement with the results of our analyis presented
in column five of Table II.  Unfortuantely, our value of
$f$ extrapolated to $T=0$  (Fig. 10) has too much uncertainty to help with
the analysis of the Fomblin stoichiometry as discussed in the
Introduction.

We hope that the analysis presented here
will stimulate high-accuracy experimental measurement of the 
temperature dependent $1/v$ cross section of Fomblin.  
It is our expectation that the temperature
dependence of $f$ together with the surface wave loss rate
can fully explain the temperature dependence of the storage
lifetime of Fomblin coated bottles.

We thank Albert Steyerl for critically reading the manuscript
and providing clarifications on several points.  SKL was supported
by LANL LDRD-DR 2001526.

\begin{table}
\caption{Results of fits to the data shown in Fig. 4.}
\begin{tabular}{||c|c|c|c|c|c||}
T (K) & q (cm$^{-1}$) & $2\pi\tau$ (s) & $y$ & $\nu$ (cm$^2$/s)
& $\alpha$ dyne/cm \\
\hline
286& 90.8& $1.3\pm 0.2\times 10^{-4}$ & $5\pm 1\times 10^{-3}$ & $2.9\pm 0.4$
& $28\pm 8$ \\
300&90.8& $2.5\pm0.2\times 10^{-4}$& $2.2\pm 0.8\times 10^{-2}$
& $1.52\pm 0.15$&$34\pm 14$ \\
309&181.6&$1.0\pm 0.1\times 10^{-4}$& $2.1\pm 0.2\times 10^{-2}$&
$0.87\pm 0.09$& $25\pm 4$  \\
347&90.8& $1.1\pm 0.1\times 10^{-3}$&$0.32\pm0.01$ & $0.30\pm 0.03$ 
& $26\pm 3$\\
\end{tabular}
\end{table}

\begin{table}
\caption{Comparison of UCN lifetime
fit results without surface wave loss
(Richardson et al, [9]) to results using
the surface wave upscattering probability
given in Sec. IIIE.  In both cases, the
number of fit parameters are the
same (two, $f$ and $A_h$).  The results from
[9] do not include the higher-energy 
lifetime data points; $\chi^2$ given in [9]
has been adjusted here to include all points (those near 100 neV
were excluded from the fit in [9]).}
\begin{tabular}{||c|c|c|c|c|c|c||}
 \multicolumn{4}{||c|}{Ref. [9] fit results}&\multicolumn{2}{l}
{With Surface Waves}& \\
\hline
Temp [K]& $f\ [10^{-5}]$ & $A_h\ {\rm [mm^2]}$ & $\chi^2$ (d.o.f.)
 & $f\ [10^{-5}]$ 
& $A_h\ {\rm [mm^2]}$ & $\chi^2$ (d.o.f.)\\
\hline
283&1.9(0.2)&3.0(0.5)&5(4)&0.83(0.10)&5.6(0.5)&1.5(5)\\  
294&2.45(0.1)&0.0(0.0)&8(8)&1.28(0.10)&3.5(0.6)&2.1(8)\\
308&3.9(0.5)&3.4(1.5)&9(2)&1.42(0.3)&8.3(1.0)&2.4(3)\\
\end{tabular}
\end{table}

\begin{table}
\caption{Recent experimental measurements of UCN upscatter probability [15]
compared with calculation.}
\begin{tabular}{||c|c|c||}
Temp (K)&Exp. Upscatter Prob.& Calc. Prob.\\
\hline
343&$9\pm 1\times 10^{-6}$&$2.4\times 10^{-5}$\\
308&$3\pm .8\times 10^{-6}$&$5.8\times 10^{-6}$\\
298&$1.2\pm .8\times 10^{-6}$&$3.4\times 10^{-6}$\\
\end{tabular}
\end{table}

\begin{figure}
\centerline{\psfig{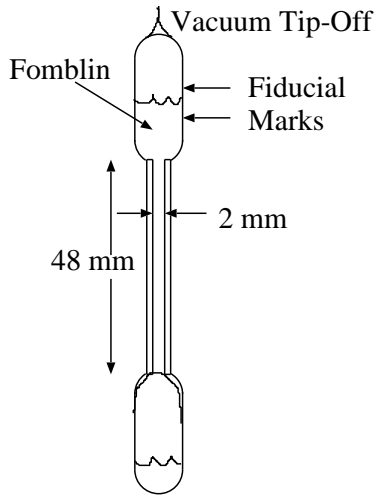}}
\caption{Schematic of the viscosity measurement apparatus.}
\end{figure}

\begin{figure}
\centerline{\psfig{figure=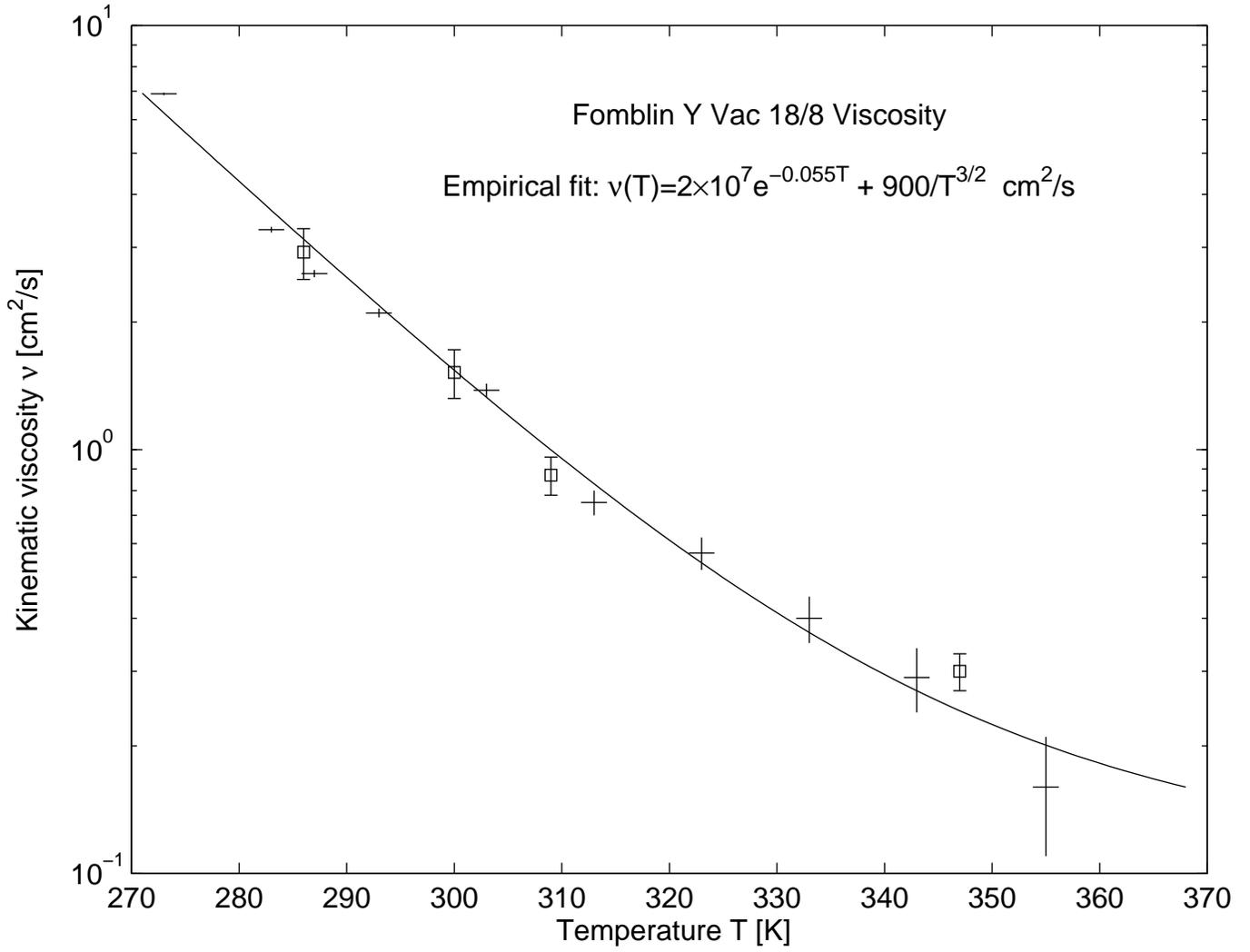}}
\caption{Temperature dependence of Fomblin viscosity from
mechanical measurements (crosses) and laser light scattering
(squares).}
\end{figure}

\begin{figure}
\centerline{\psfig{figure=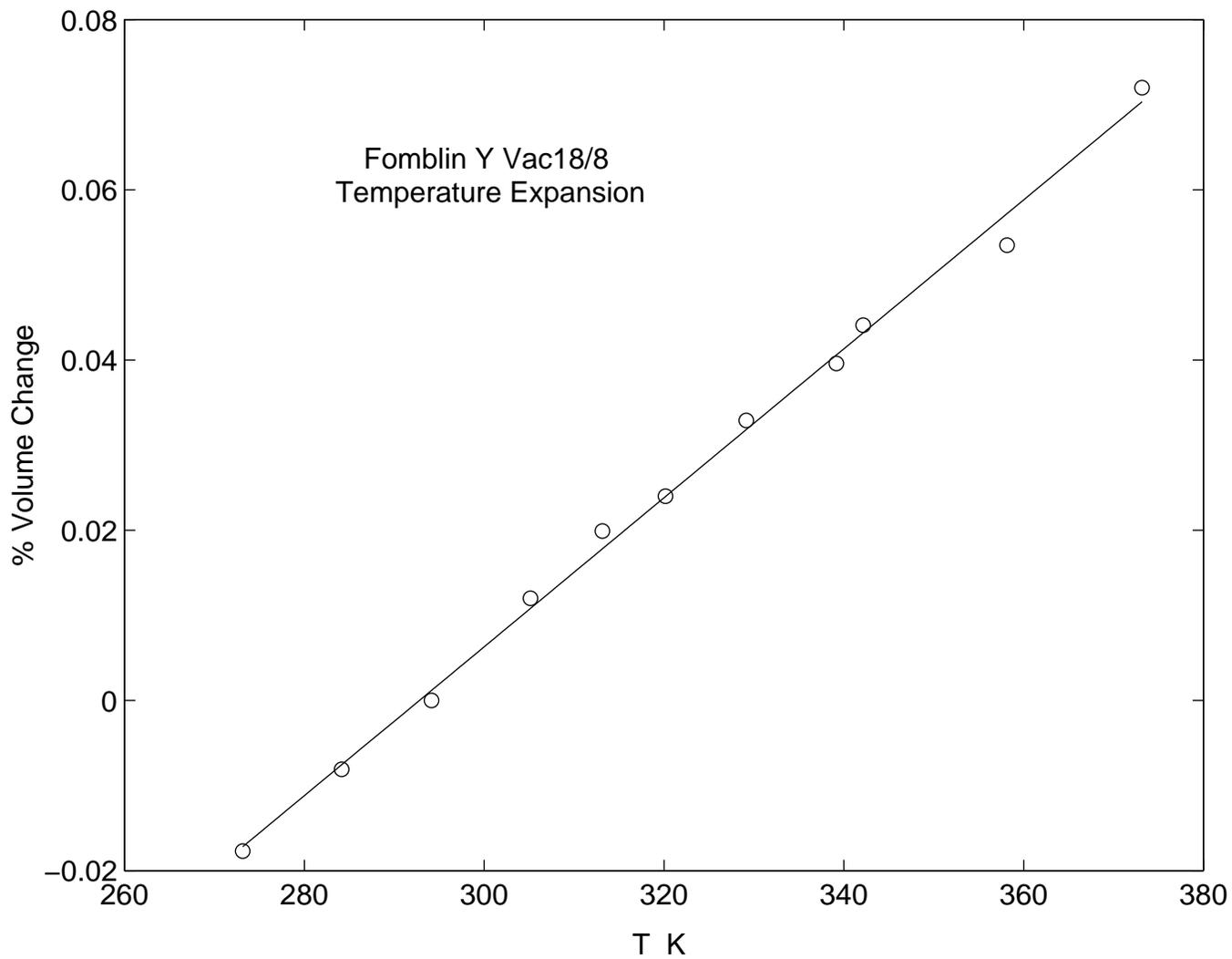}}
\caption{Volume expansion of Fomblin with temperature.}
\end{figure}

\begin{figure}
\centerline{\psfig{figure=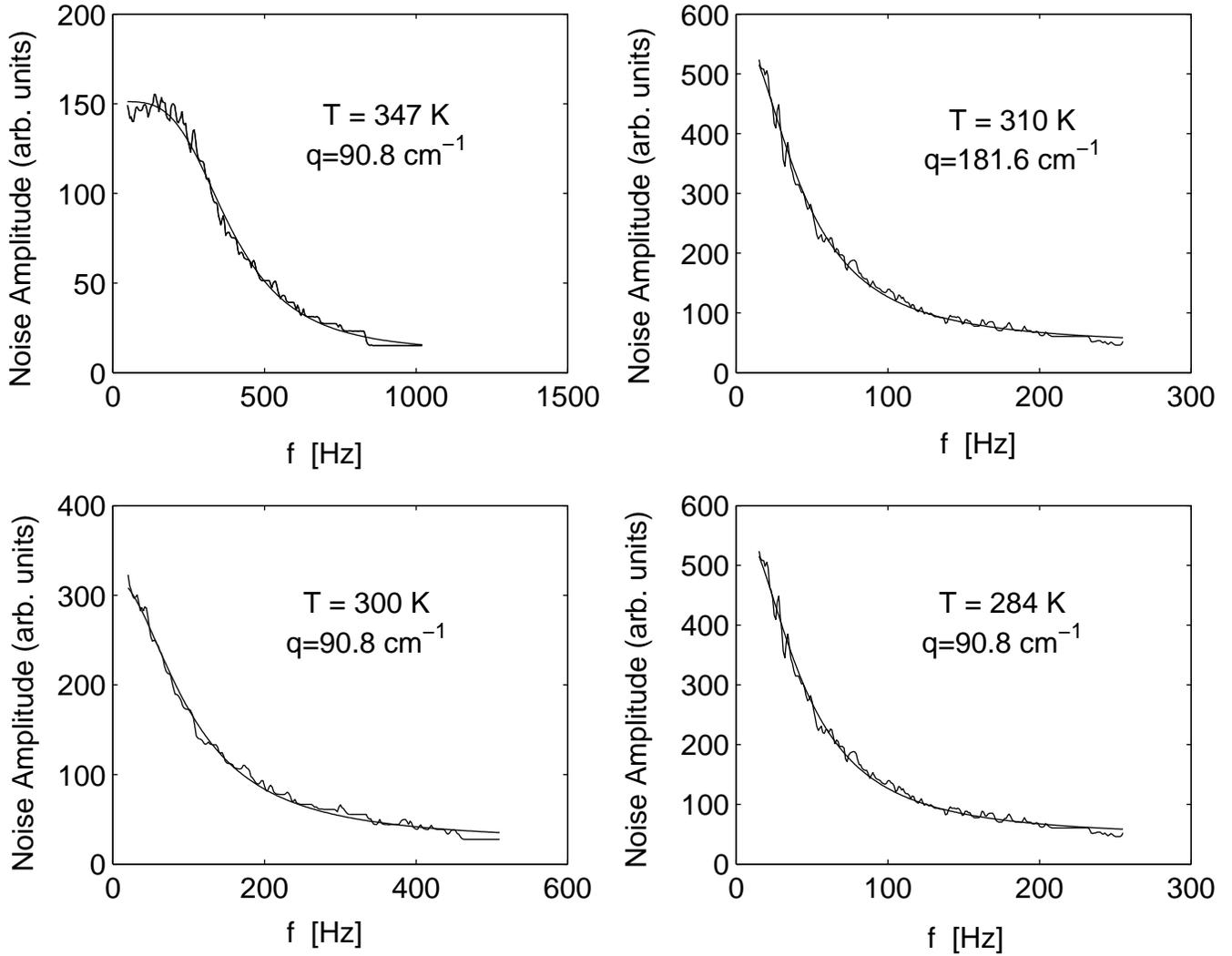}}
\caption{Experimental measurements of heterodyne detected 
inelastic light scattering with theoretical fits, for various
temperatures as indicated in plots.}
\end{figure}

\begin{figure}
\centerline{\psfig{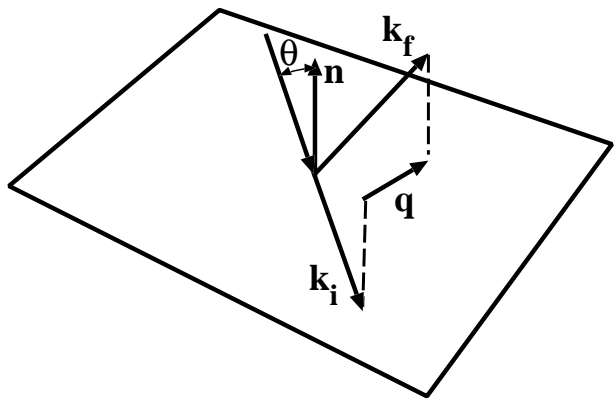}}
\caption{Kinematics of UCN scattering from surface waves; final energy
greater than initial energy.}
\end{figure}

\begin{figure}
\centerline{\psfig{figure=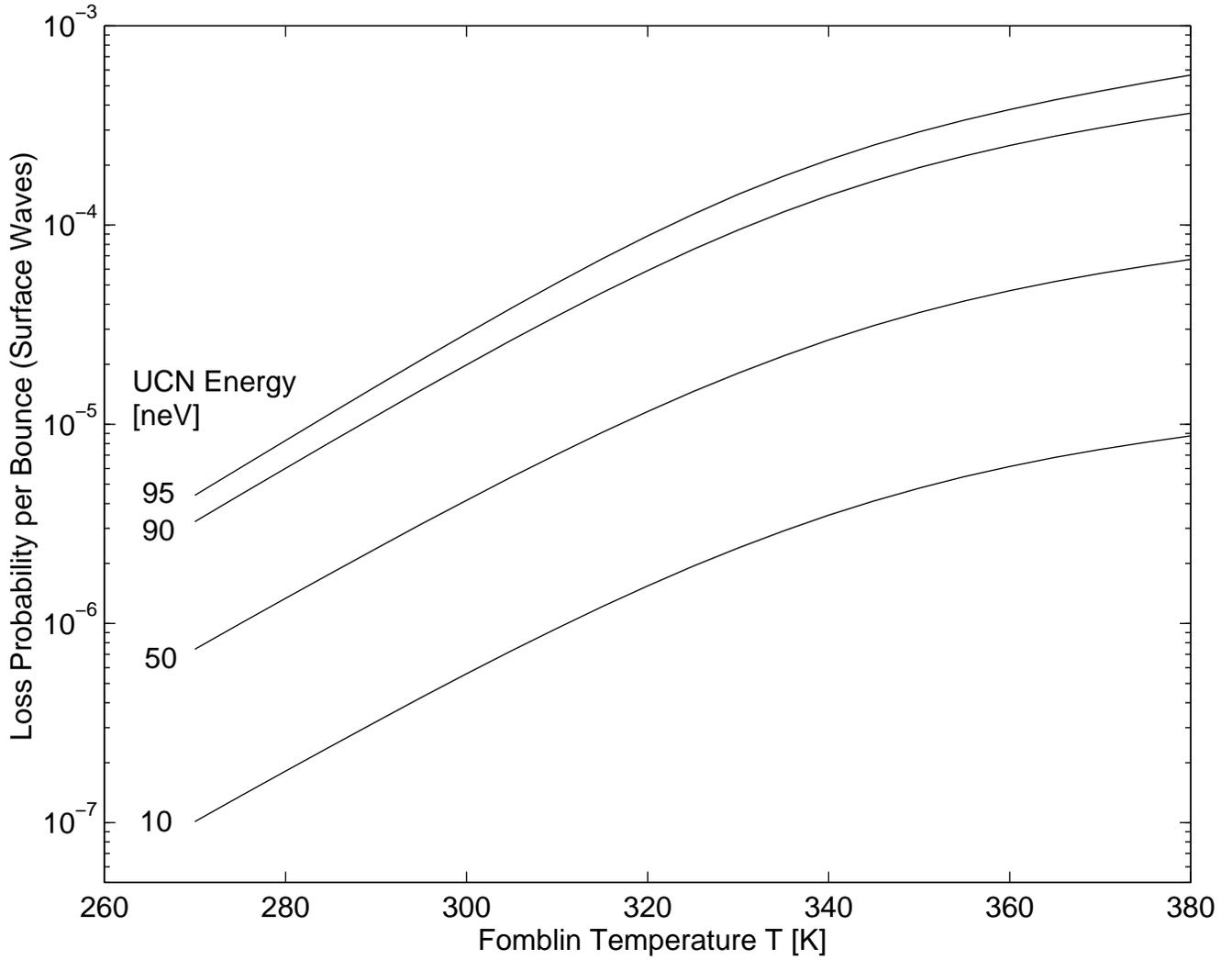}}
\caption{Probability for UCN to upscatter to energy higher than the
wall potential, as a function of temperature.  Included are the
temperature dependencies of the viscosity and of the density. The
effective potential is modified for the density change.}
\end{figure}

\begin{figure}
\centerline{\psfig{figure=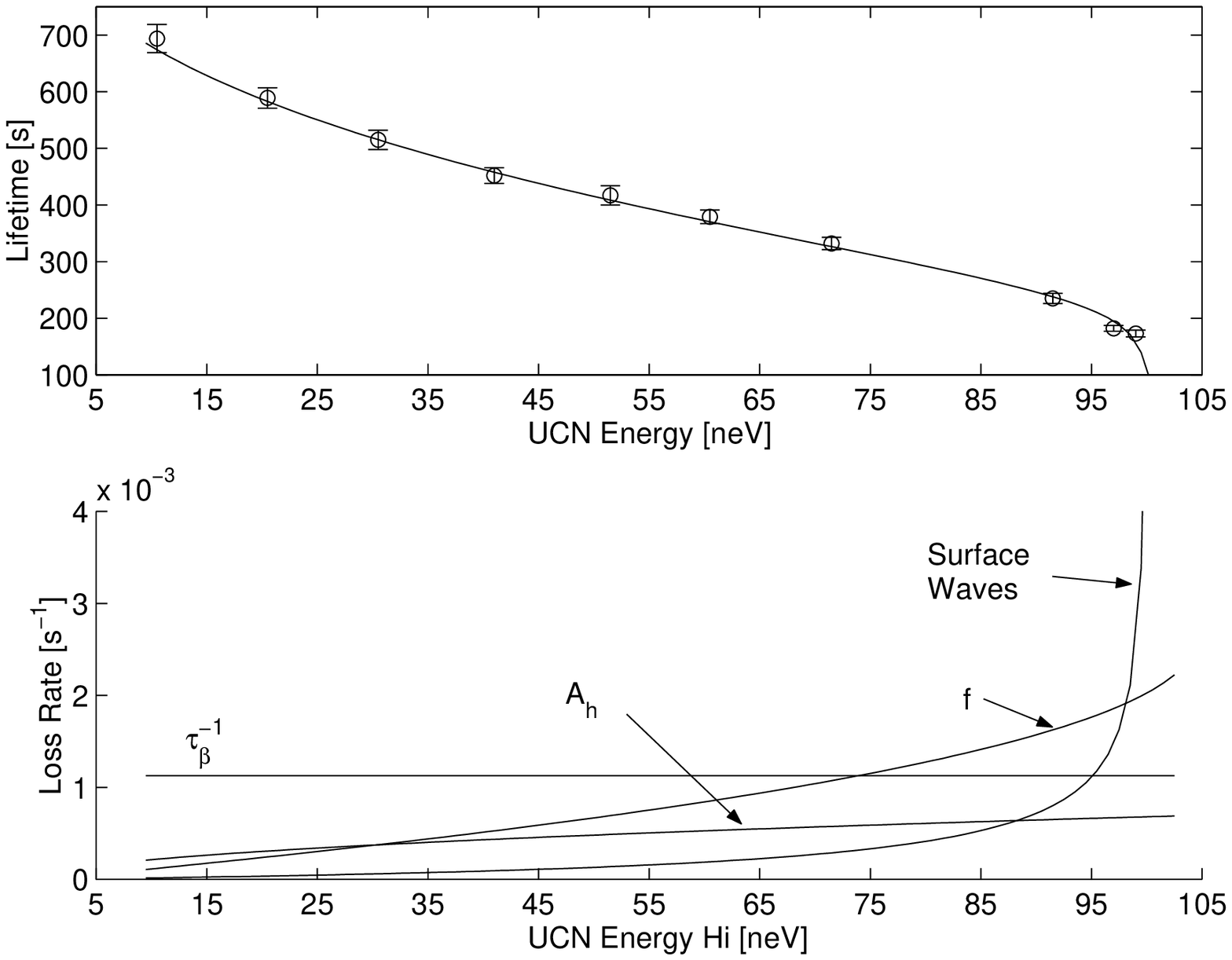}}
\caption{Least square fit to 294 K data from [11]. The experimentally
determined values $\rho=1.89$ g/cc,
$\nu=2.0 {\rm cm^2/s}$, and $\alpha=24$ dyne/cm$^2$
were used in the
fit to $A_h$ and $f$, resulting in a  reduced $\chi^2=2.1$ for the fit. Upper
plot shows the net lifetime, while the lower plot shows the contributions
from the various loss mechanisms.}
\end{figure}

\begin{figure}
\centerline{\psfig{figure=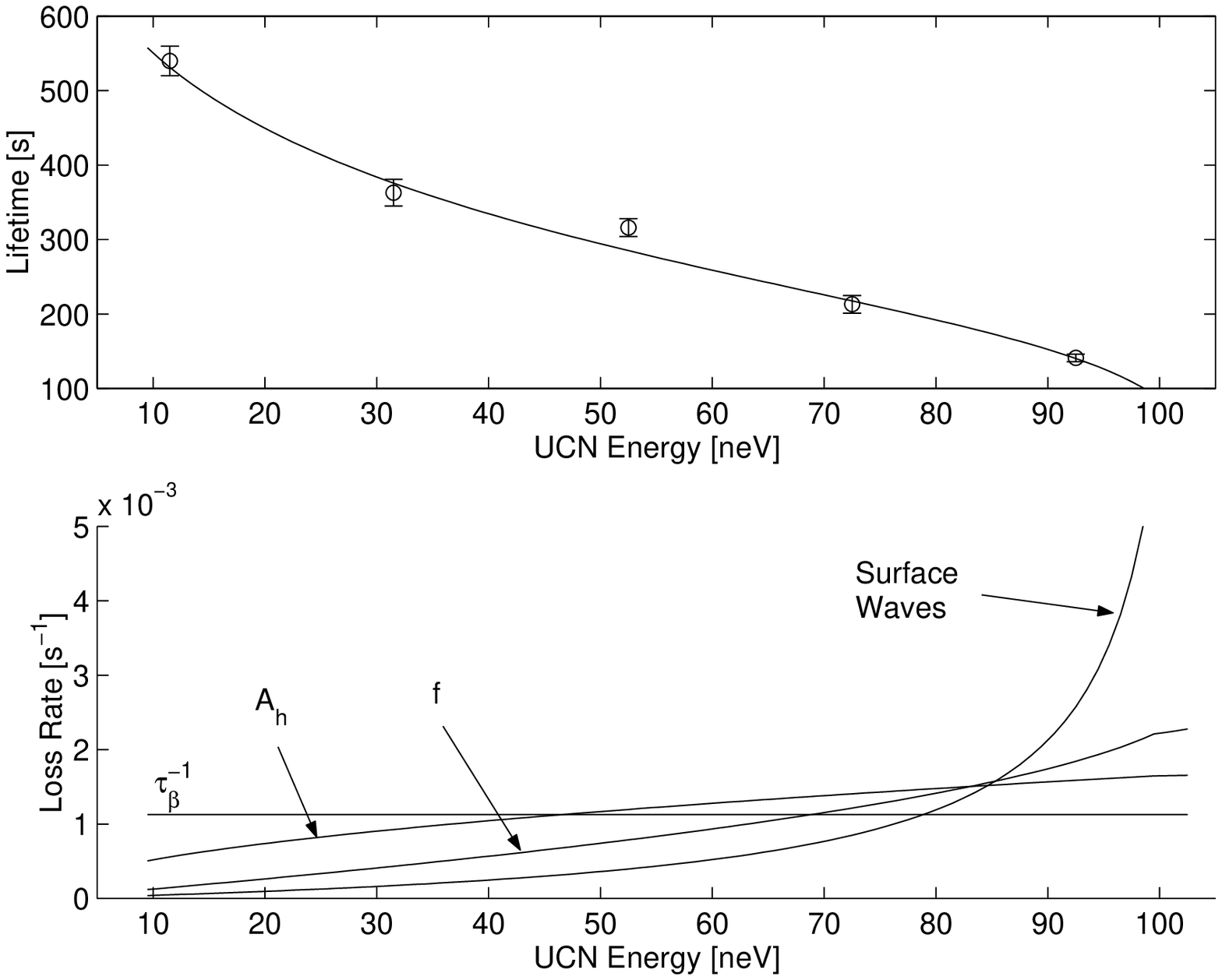}}
\caption{Least square fit to 308 K data from [11]. The experimentally
determined values $\rho=1.87$ g/cc,
$\nu=1.0 {\rm cm^2/s}$, and $\alpha=24$ dyne/cm$^2$ 
were used in the fit to $A_h$ and $f$, resulting
in a reduced $\chi^2=2.4$ for the fit. Upper
plot shows the net lifetime, while the lower plot shows the
contributions to the loss rate from the various mechanisms.}
\end{figure}

\begin{figure}
\centerline{\psfig{figure=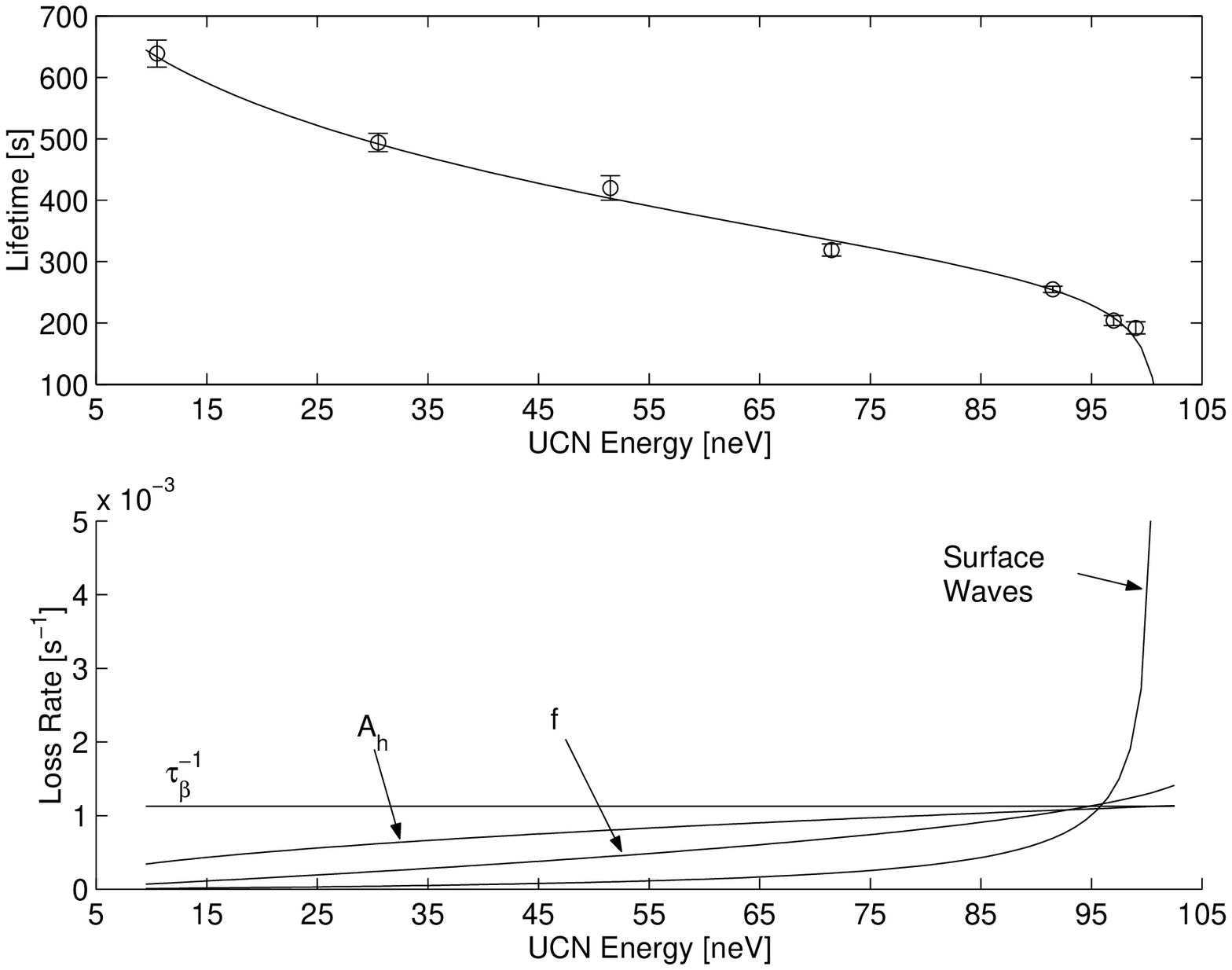}}
\caption{Least square fit to 283 K data from [11].  The experimentally
determined values $\rho=1.91$ g/cc,
$\nu=3.0$ cm$^2$/s, and $\alpha=24$ dyne/cm$^2$= were used in the fit
to $A_h$ and $f$, 
resulting in a reduced $\chi^2=1.5$ for the fit. Upper plot shows
the net lifetime, while the lower plot shows the contribution
to the loss rate from the various mechanisms. }
\end{figure}

\begin{figure}
\centerline{\psfig{figure=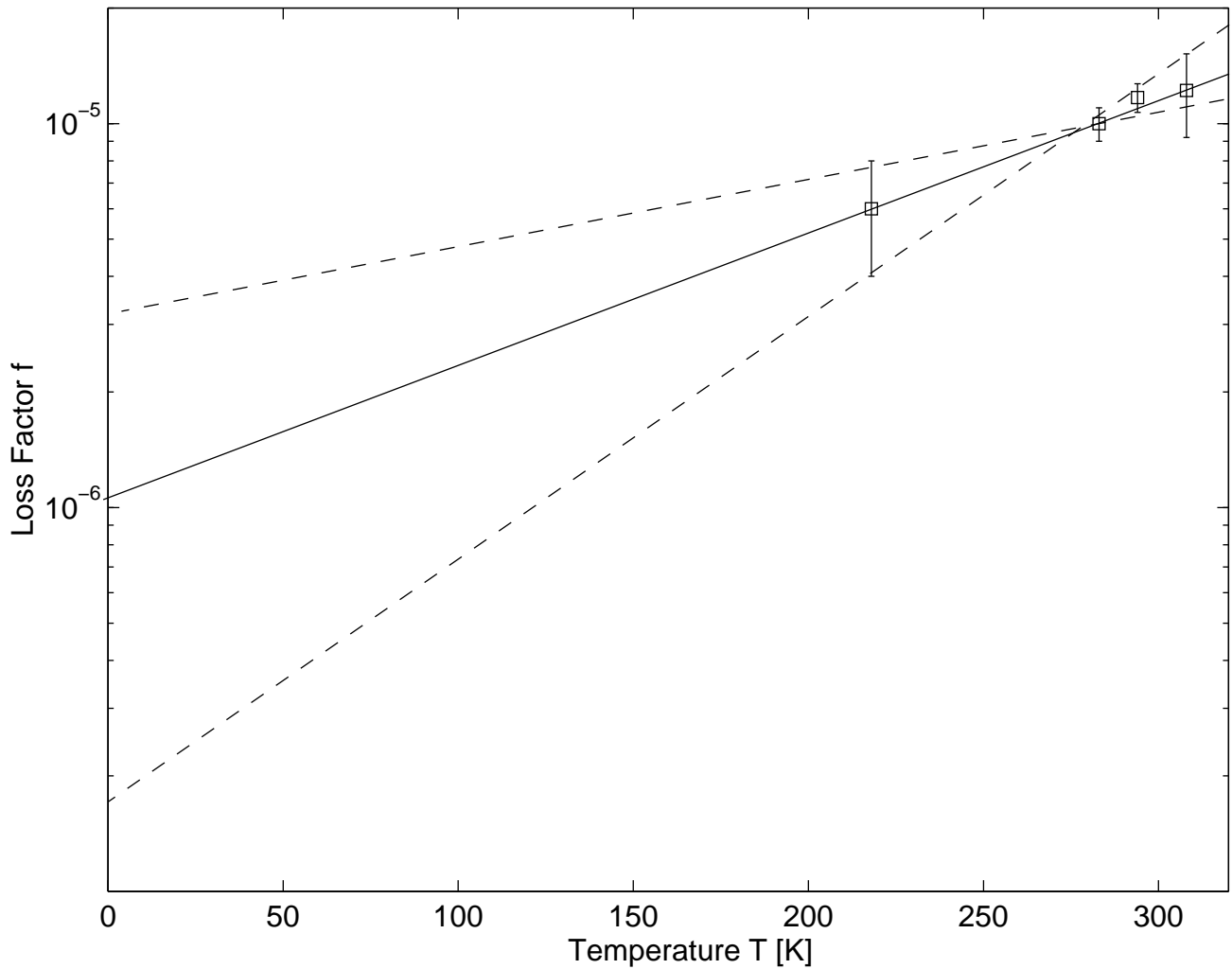}}
\caption{Loss factor from the surface wave least-square
fits (column five in Table II) and
the 218 K value derived from [7] plotted as a function
of temperature.  A simple extrapolation to $T=0$ yields
a range of $f$ consistent with
the  nuclear absorption value.}
\end{figure}

\begin{figure}
\centerline{\psfig{figure=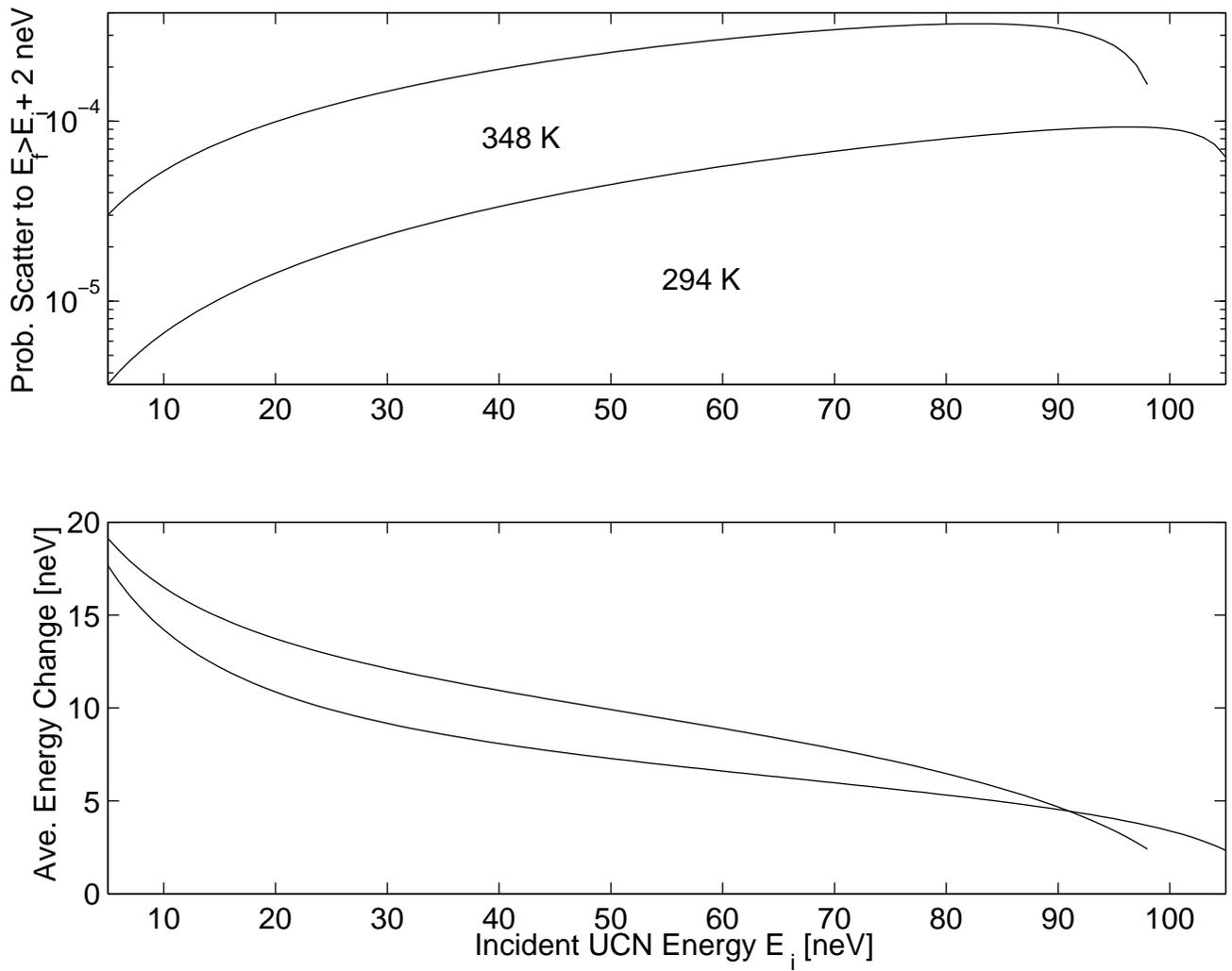}}
\caption{Upscatter heating probability, with final energy
2 neV greater than the initial energy.  Lower plot, the average change
in energy with upscattering.}
\end{figure}

\begin{figure}
\centerline{\psfig{figure=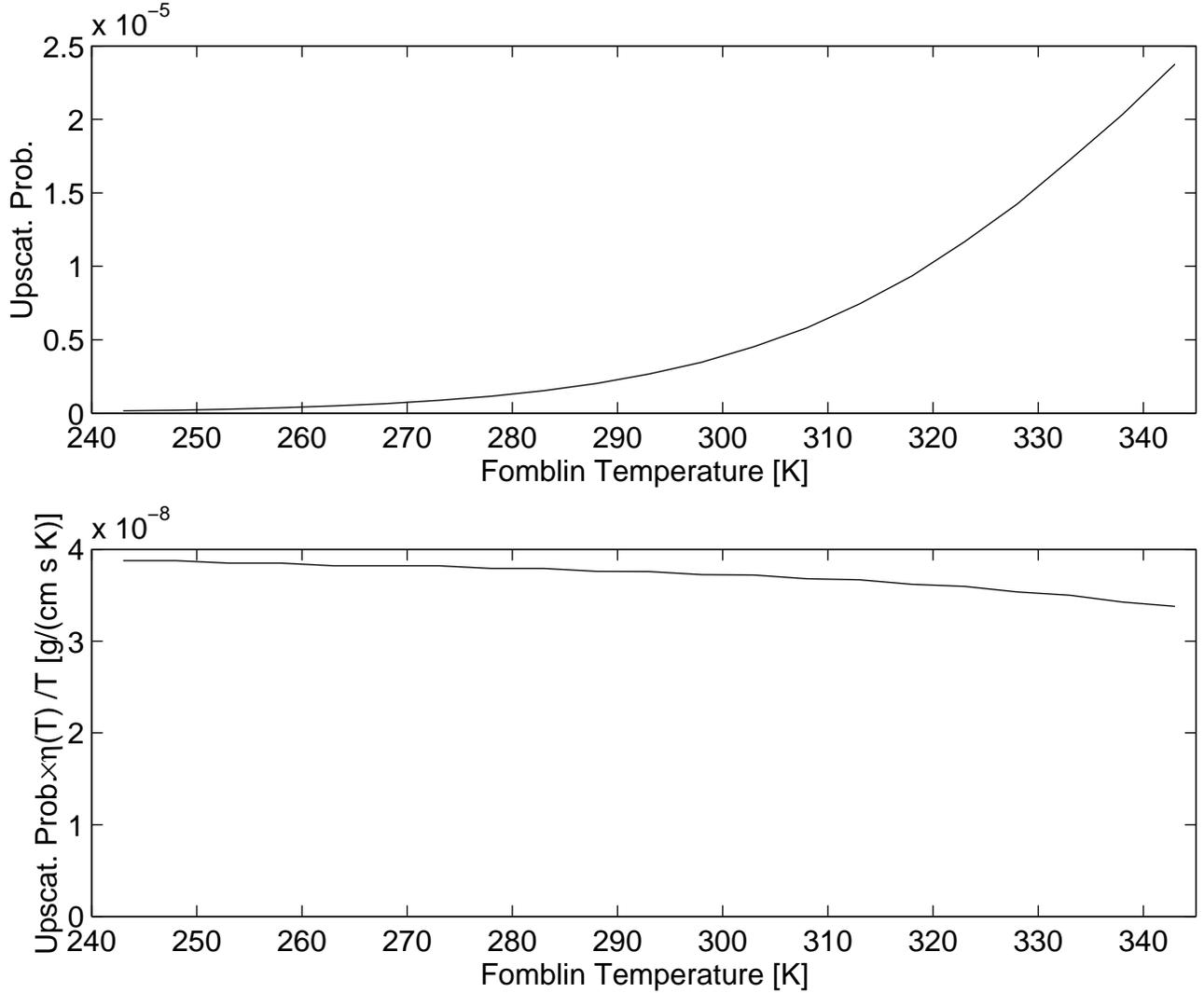}}
\caption{Upper Plot: The upscatter probability per bounce
for UCN  in the energy range
0-52 neV range, with initial distribution given in Fig. 5 of [17],
to upscatter to the 52-106 neV energy range, as a function
of Fomblin temperature. Lower Plot: The upscatter
probability times the viscosity, divided by temperature, which
is approximately constant for $T<300$ K.  This is because in
this temperature range the
$y$ term in Eq. (8) is relatively unimportant, and $S<<1$.
This leaves Eq. (8) proportional to $T/\eta(T)$, with the integration
over the $q$ and $\omega$ factors contributing an overall factor
roughly independent of $T$.}
\end{figure}

\begin{figure}
\centerline{\psfig{figure=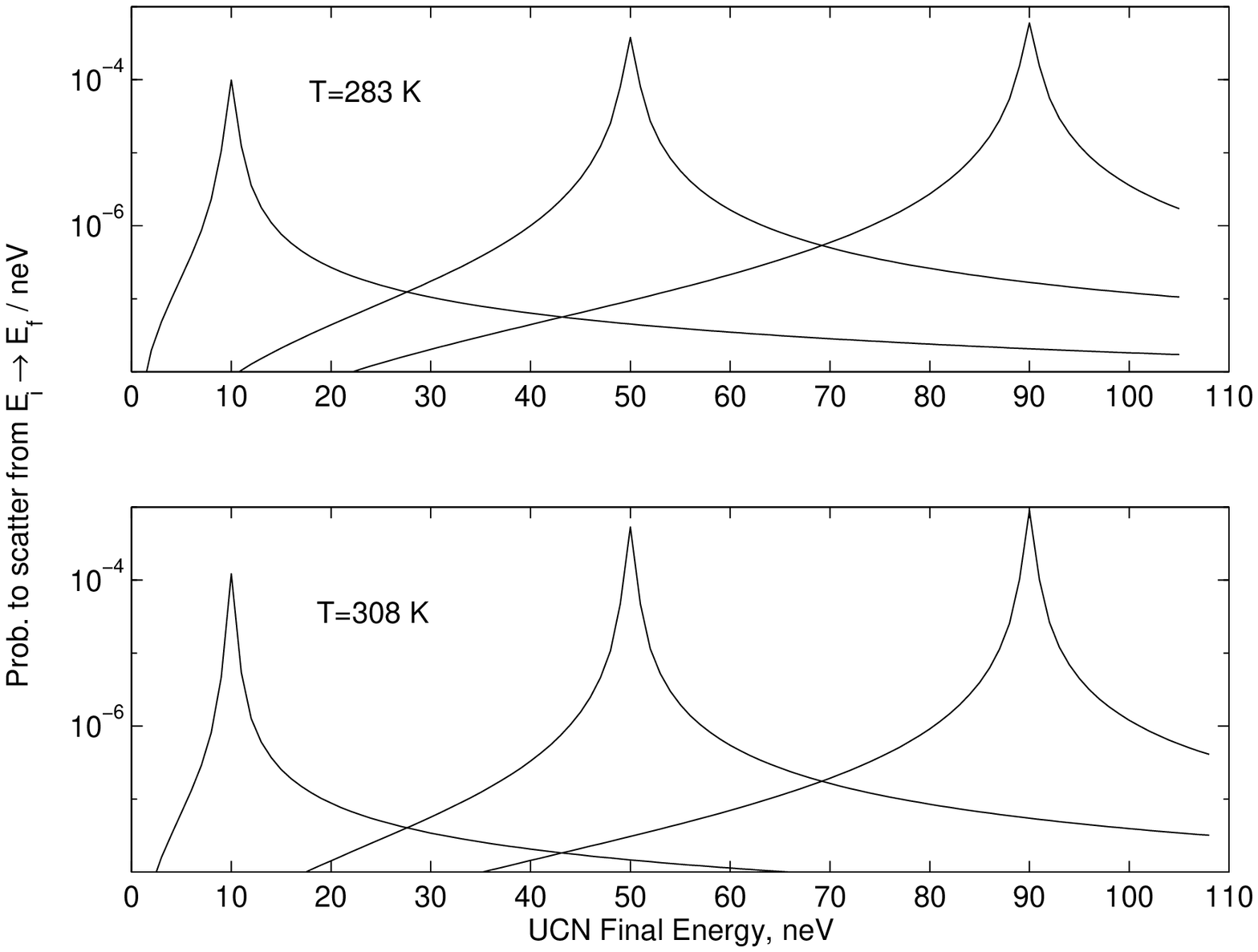}}
\caption{The probability for UCN to up- or down-scatter,
per neV final energy, for three different UCN initial energies.
These results are similar to those presented in [9,10].}
\end{figure}


\begin{thebibliography}{99}

\bibitem{bates} J.C. Bates, Nucl. Instr. and Meth. {\bf A249}, 261 (1986).

\bibitem{mampe} W. Mampe et al, Nucl. Instr. and Meth. {\bf A284}, 111 (1989);
Phys. Rev. Lett. {\bf 63}, 593 (1989).

\bibitem{arzu} S. Arzumanov et al., Phys. Lett. B {\bf 483}, 15 (2000).


\bibitem{ucnbook} {\it Ultracold Neutrons}, R. Golub, D. Richardson,
S.K. Lamoreaux (Adam-Hilger, Bristol, 1991).

\bibitem{fomform} F. Tervisidis and N. Tsaga, Nucl. Instr. and Meth.
{\bf A305}, 433 (1991).

\bibitem{ruddies} R. Ruddies, W. Mampe, D. Dubbers, 
Institut Laue-Langevin Praktikum ``Messung der temperaturah\"angigen
Streuung von UCN an Fomblin\"ol,'' 1988 (unpublished). 

\bibitem{moroz} W. Mampe et al., JETP Letters (Eng. Trans.)
{\bf 57}, 82 (1993).

\bibitem{moroz2} V.I. Morozov, Private Communication, 2002.

\bibitem{poko1} Yu. N. Pokotilovski, Eur. Phys. Jour. {\bf B8}, 1 (1999).

\bibitem{poko2} Yu. N. Pokotilovski, Phys. Lett. {\bf A255}, 173 (1999).

\bibitem{dave} D.J. Richardson et al., Nucl. Instr. and Meth. {\bf A308},
568 (1991).

\bibitem{ll} L.D. Landau and E.M. Lifshitz, {\it Fluid Mechanics}
(Pergamon, Oxford, 1987).

\bibitem{bom} M.A. Bouchiat and J. Meunier, J. de Phys. {\bf 32}, 561 (1972).

\bibitem{hard} S. H\aa rd, Y. Hamnerius, and O. Nilsson, Appl. Phys.
{\bf 47}, 2433 (1976).

\bibitem{ajp} W.M. Klipstein, J.S. Radnich, and S.K. Lamoreaux,
Amer. Jour. Phys. {\bf 64}, 768 (1996).

\bibitem{asty} Albert Steyerl, Private Communication, 2002.

\bibitem{moroz3} L.N. Bondarenko et al., Phys. of Atom. Nuclei {\bf 65},
11 (2002). (Trans. from Yad. Fiz. {\bf 65}, 13 (2002))

\end{thebibliography}
\end{document}